\documentclass[amssymb,useAMS,prd,aps,amsmath,twocolumn,superscriptaddress,nofootinbib,preprintnumbers]{revtex4-1}
\usepackage{graphicx}
\usepackage{xcolor}
\usepackage{placeins}
\usepackage{ulem}
\newcommand{\remove}[1]{}

\begin{document}

\preprint{FERMILAB-PUB-20-247-T}
\preprint{IFT-UAM/CSIC-20-88}
 
\title{Light new physics in XENON1T}

\author{C\'eline B\oe hm} 
\email{celine.boehm@sydney.edu.au}
\affiliation{School of Physics, Physics Road, The University of Sydney, NSW 2006 Camperdown, Sydney, Australia}

\author{David G. Cerde\~no}
\email{davidg.cerdeno@gmail.com}
\affiliation{Instituto de F\' isica Te\'orica, Universidad Aut\'onoma de Madrid, 28049 Madrid, Spain}

\author{Malcolm Fairbairn}
\email{malcolm.fairbairn@kcl.ac.uk}
\affiliation{Physics, King's College London, Strand, London WC2R 2LS, UK}

\author{Pedro A. N. Machado}
\email{pmachado@fnal.gov}
\affiliation{Theory Department, Fermi National Accelerator Laboratory, P.O. Box 500, Batavia, IL 60510, USA}

\author{Aaron C. Vincent}
\email{aaron.vincent@queensu.ca}
\affiliation{Department of Physics, Engineering Physics and Astronomy, Queen's University, Kingston ON K7L 3N6, Canada}
\affiliation{Arthur B. McDonald Canadian Astroparticle Physics Research Institute, Kingston ON K7L 3N6, Canada}
\affiliation{Perimeter Institute for Theoretical Physics, Waterloo ON N2L 2Y5, Canada}

\begin{abstract}
We examine the recently-reported low-energy electron recoil spectrum observed at the XENON1T underground dark matter direct detection experiment, in the context of new interactions with solar neutrinos. In particular we show that scalar and vector mediators with masses $\lesssim 50$ keV coupled to leptons could already leave a visible signature in the XENON1T experiment, similar to the observed peak below 7 keV. This signals that dark matter detectors are already competing with neutrino scattering experiments such as GEMMA, CHARM-II and Borexino.  If these results from XENON1T are interpreted as a new signal of such physics, the parameters which fit the excess face challenges from astrophysics which seem very difficult to overcome.  If they are rather viewed as a constraint on new couplings, they herald the start of an era of novel precise probes of physics beyond the standard model with dark matter detectors.
\end{abstract}
\maketitle

\section{Introduction}
Underground direct detection experiments were proposed and developed to probe the existence of weakly-interacting dark matter particles. While they were initially designed to probe a dark matter particle mass range above a few GeV \cite{Goodman:1984dc},  models of thermal sub-GeV dark matter particles \cite{Boehm:2002yz,Boehm:2003hm} together with the detection of mysterious cosmic ray excesses \cite{Jean:2003ci,Adriani:2008zr,Jeltema:2014qfa} and the absence of a confirmed dark matter signal have led to a shift of paradigm in the community \cite{Boehm:2003bt,Gunion:2005rw,Pospelov:2007mp,Huh:2007zw,Finkbeiner:2007kk,Gelmini:2008fq,Feng:2008dz} and encouraged experimental collaborations to search for light dark matter particles and light mediators (such as the ones proposed in \cite{Fayet:1977yc,Boehm:2003hm}).   

Currently, the strongest limits on spin-independent WIMP dark matter above {10}~GeV are from the  XENON1T experiment \cite{Aprile:2017aty,Aprile:2018dbl}, a liquid xenon time projection chamber with an active target of 2 tonnes that operated at the Laboratori Nazionali del Gran Sasso between 2016 and 2018. XENON1T primary goal was to look for nuclear recoils (NRs) in the 1-100 keV range. 
The dual phase TPC is able to discriminate energy deposition by nuclear recoils from electron recoils normally caused by background gamma ray events. Nonetheless, low momentum transfer processes are more easily visible, i.e., lead to larger recoil energies, in the electron recoil channel due to the small electron mass compared to nuclear masses.
Thanks to low levels of external radiation and exquisitely-modeled backgrounds, the experiment is now sensitive to electronic recoil energy down to around a keV, with a higher exposure than other experiments. Compellingly, they have recently presented results from the analysis of the low-energy electronic recoil (ER) spectrum obtained in their Science Run 1, which features a 3.5 $\sigma$ excess of events in the 1-7 keV region with 285  ER events, well above the expected 232$\pm$15. The analysis rules  out backgrounds from known particles and potential systematic effects  \cite{Aprile:2020tmw}. While the collaboration could not exclude a potential contamination from tritium at this time, they interpreted the excess as a possible sign of new physics. In particular they suggested that solar axions (as initially suggested in \cite{Weinberg:1977ma,Wilczek:1977pj}) or a larger neutrino dipole moment  \cite{PhysRevLett.95.151802,Bell:2006wi} could lead to a similar signature in the recoil energy spectrum. Ref \cite{Takahashi:2020bpq} has proposed a model of dark matter axions, and Ref.~\cite{Kannike:2020agf} suggested that mildly relativistic light dark matter could also leave such an imprint, though exotic physics would be required to produce such a population.
Models in which $3\to2$ processes (e.g. $\chi+\chi+\chi\to\chi+\chi$ processes where $\chi$ denotes a dark matter particle) involving dark matter are dominant could also lead to a bump in the electron recoil spectrum at lower energies~\cite{Smirnov:2020zwf}.

In Ref.~\cite{Cerdeno:2016sfi}, expanding on the results of \cite{Pospelov:2011ha, Pospelov:2012gm, Harnik:2012ni}, we demonstrated that the next generation of liquid xenon experiments could detect new physics in the neutrino sector; a milestone considering that Direct Detection (DD) detectors were originally built to probe the nature of dark matter. 
Our analysis was focusing on the signatures that scalar and vector particles (mediators of the interactions between the visible and dark sectors) would have on the number of solar neutrinos hitting DD experiments. We showed in particular that such light particles would enhance the electron recoil rates at low  energy, thus allowing the next generation of dark matter direct detection experiments to probe their existence. 

Here we repeat this analysis in light of the new XENON1T measurements and determine the value of the couplings that would best fit the signal.  Our hypothesis is that the signal originates from the interaction of solar neutrinos with electrons in the xenon target via some new light mediator beyond the Standard Model of particle physics.  Should the signal originate from tritium contamination or another contaminant such as argon, our region would indicate that XENON1T is now sensitive to mediator couplings as small as a few $10^{-7}$, which is around three times better than the limit set by dedicated neutrino experiments such as NuTeV, CHARM-II and GEMMA \cite{Boehm:2004uq, Cerdeno:2016sfi}.  

In the next section we will set out the interactions and cross sections that we will be using in this work to attempt to fit the anomaly.  We will then describe the constraints which exist on such new couplings from other experiments and also from astrophysics.  Following that in the results section we will present the parameters which best fit the anomaly.  Finally we will conclude and make some comments.

\section{Analysis}

We consider the flux of $pp$ neutrinos from the Sun and their interaction with the electrons of the XENON1T experiment in new physics scenarios. In the energy range considered here, neutrino fluxes from other processes such as $^7$Be or $^8$B decay do not contribute meaningfully. In particular, we investigate how light mediators can give rise to more recoil events in the detector.

The number of recoils in the detector per unit energy can be written as 
\begin{equation}
\frac{dR}{dE_R} =  \frac{\epsilon}{m_T} \int dE_\nu \frac{d\phi_{\nu }}{dE_\nu} \frac{d \sigma_{\nu }}{dE_R},
\end{equation}
where $\epsilon$ is the exposure and $m_{T}$ is the mass of the target electron or nucleus. 

In the Standard Model, the neutrino-electron scattering cross section is given by 
\begin{eqnarray}
\frac{d \sigma^{SM}_{\nu e}}{dE_R} &=& \mathcal{F}(E_R)\frac{G_F^2 m_e}{2 \pi}\biggl[ (g_v + g_a)^2  + \label{eq:SMxsec} \\
&& (g_v-g_a)^2\left(1-\frac{E_R}{E_\nu}\right)^2 + (g_a^2 - g_v^2)\frac{m_eE_R}{E_\nu^2} \biggr], \nonumber
\end{eqnarray}
where $G_F$ is the Fermi constant, $m_e$ is the electron mass, $E_R$ is the electron recoil energy and $E_\nu$ is the incoming neutrino energy. 
The $g_v$ and $g_a$ couplings depend on the neutrino flavor. For electron neutrinos we have
\begin{equation}\label{eq:gv_ga_e}
g_{v}^e = 2\sin^2 \theta_W + \frac{1}{2}; \, \, \, g_{a}^e  = +\frac{1}{2}, 
\end{equation} 
while for muon and tau neutrinos
\begin{equation}\label{eq:gv_ga_mu}
g_{v}^{\mu,\tau} = 2\sin^2 \theta_W - \frac{1}{2}; \, \, \, g_{a}^{\mu,\tau}  = -\frac{1}{2}, 
\end{equation} 
where $\sin^2\theta_W=0.23$ \cite{Tanabashi:2018oca} is the weak mixing angle. 
Neutrinos coming from the Sun are an incoherent admixture of $\nu_e$ and $\nu_{\mu,\tau}$.
The $P(\nu_e\to\nu_e)$ survival probability inferred from solar neutrino measurements is approximately 55\% \cite{Aharmim:2011vm, Abe:2016nxk, Agostini:2018uly}.
To avoid clutter, we will omit the flavor indices from $g_v$ and $g_a$ in the cross section formulas that will be discussed below.

Eq.~\eqref{eq:SMxsec} changes in the presence of new mediators of the interactions between neutrino $\nu$ and electron  $e$. In the following we will consider three types of new mediators, namely 
a spin-0 particle with scalar couplings, a spin-1 particle with axial couplings and a spin-1 particle with vector  couplings. 
In principle we could also have a pseudoscalar mediator, but since the cross section is not enhanced at low recoils (see Ref.~\cite{Cerdeno:2016sfi}), constraints from higher energy experiments that have measured $\nu-e$ scattering cross section will typically rule out this scenario. We have verified that this scenario does not lead to an improvement in either fit or exclusions.
Therefore, we do not consider it here.
The expected deviation from the Standard Model $\nu-e$ scattering cross section for each of these cases are summarised below. Note that the momentum transfer is given by $q^2=-2E_Rm_e$, where $E_R$ and $m_e$ are the recoil energy and mass of the electron.

\begin{itemize}
\item Scalar mediator (with mass $m_\phi$) and scalar couplings. The effective Lagrangian is

\begin{equation}
\mathcal{L}\supset\left(g_{\nu}\,\phi \bar{\nu}_R \nu_L + h.c.\right) \newline + g_{e}\phi\, \bar{e}  e
\end{equation}
($\phi$ is the new scalar, $\nu$ and $e$ are the standard model neutrino and electron fields, $g_\nu$ and $g_e$ are coupling constants), leading to the differential cross section
\begin{equation}\label{eq:scalar-xsec}
    \frac{d\sigma_{\nu e}}{dE_R}-\frac{d\sigma^{\rm SM}_{\nu e}}{dE_R}=\frac{g_{\nu}^2 g_{e}^2 E_R m_e^2 }
      {4 \pi E_{\nu }^2 \left(2 E_R m_e+m_\phi^2\right)^2}~.
      \end{equation}

\item Vector mediator $Z'$ (with mass $m_{Z'}$) and vector couplings. The effective Lagrangian is
\begin{equation}
\mathcal{L}=g_{\nu} Z'_\mu \bar{\nu}_L \gamma^\mu \nu_L+ g_{e} Z'_\mu \bar{e} \gamma^\mu   e 
\end{equation}
($Z'$ is the new vector boson), leading to the differential cross section
 \begin{eqnarray}\label{eq:vector-xsec}
 \frac{d\sigma_{\nu e}}{dE_R}-\frac{d\sigma^{\rm SM}_{\nu e}}{dE_R}=&&\frac{\sqrt{2}G_Fm_e g_vg_{\nu}g_{e}}{\pi\left(2 E_R m_e+m_{Z'}^2\right)} \nonumber\\
  &+&\frac{m_e g_{\nu}^2g_{e}^2}{2\pi\left(2 E_R m_e+m_{Z'}^2\right)^2}~.
  \end{eqnarray}

\item Vector mediator $Z'$ (with mass $m_{Z'}$) and axial vector couplings. The effective Lagrangian is
\begin{equation}
\mathcal{L}=g_{\nu,Z'} Z'_\mu \bar{\nu}_L \gamma^\mu \nu_L \newline - g_{e} Z'_\mu \bar{e} \gamma^\mu \gamma^5 e, \newline
\end{equation}
leading to the differential cross section
\begin{eqnarray}
  \frac{d\sigma_{\nu e}}{dE_R}-\frac{d\sigma^{\rm SM}_{\nu e}}{dE_R}&=&\frac{\sqrt{2}G_Fm_e g_a g_{e}g_{\nu}}{\pi\left(2 E_R m_e+m_{Z'}^2\right)}\nonumber\\
  &+&\frac{m_e g_{\nu}^2g_{e}^2}{2\pi\left(2 E_R m_e+m_{Z'}^2\right)^2}~.
  \end{eqnarray}

\end{itemize}
We note that due to the structure of the couplings, there can be a positive or negative interference between the vector or axial amplitudes and the standard model amplitude, depending on the values of $g_e$ and $g_\nu$. For simplicity, we assume $g_e$ and $g_\nu$ to have the same sign throughout this paper.

Such new forces are already tightly constrained, since new (light) mediators induce new electron-electron, electron-neutrino and neutrino-neutrino interactions.
For example, constraints on the coupling of neutrinos to each other via a new mediator have been studied extensively in the context of majoron theories.  In particular, if the self-coupling of neutrinos is too large, then supernovae cores will cool too rapidly, resulting in a lower overall luminosity of neutrinos, even if the mediators decay back into $\nu$ \cite{Heurtier:2016otg,Farzan:2002wx}.  This leads to a constraint on the coupling of mediators to a new scalar which are approximately
\begin{equation}
g_{\nu}\lesssim 2\times 10^{-9}\left(\frac{\rm MeV}{m_{\phi}}\right).
\label{eq:SNbound}
\end{equation}

The constraints on $g_e$ from $e-e$ scattering in supernovae, red giants and the Sun are even more constraining across the entire region of interest as we will discuss in Sec. \ref{sec:conclusions}. 
 There is however a strong constraint from the effect that new energy transport mechanisms can have upon both the Sun and hotter helium burning stars observed in globular clusters, which yields very strong constraints on new interactions between electrons due to the presence of a new mediator.  Naively it seems that any mass less than around 0.1 MeV would be ruled out \cite{An:2013yfc,Redondo:2013lna} however as pointed out by the authors of \cite{An:2013yfc} it is possible that islands exist where interactions between new mediators and electrons are sufficiently strong as to prevent deviation from normal energy loss or normal energy transport within the star becoming important.  Previous analyses, for example \cite{Harnik:2012ni} have pointed out that energy loss of a mediator with couplings above $10^{-8}$ can be considerable over a short distance.  Further study seems to be required to shed light on the possibility of such interactions being viable within stars.   For the time being we will proceed without recourse to the effects of these astrophysical constraints on the acceptable parameter space and discuss them at the end of the paper.

In terms of constraints on new electron-neutrino couplings the GEMMA experiment \cite{Beda:2013mta} yields a strong constraint.
Other experiments also contribute significantly to these constraints, such as TEXONO~\cite{Deniz:2010mp} (low masses), Borexino~\cite{Harnik:2012ni, Bilmis:2015lja, Agostini:2018uly} (intermediate masses) and CHARM-II (higher masses)~\cite{Boehm:2004uq}. In the region of interest for us, GEMMA and Borexino provide the dominant constraints.
We recast the GEMMA constraint on neutrino magnetic moment into scalar mediators by simply equating the differential cross sections of magnetic moment and Eq.~\eqref{eq:scalar-xsec} at 3~keV electron recoil and average reactor neutrino energy of 500~keV.

New electron-electron interactions should be sensitive to the presence of a fifth force for mediator masses below 100 eV and atomic physics constraints are important at keV masses and much larger couplings, but neither regime will be important for the regions favoured by the fit.  

Another constraint is the effect of a new gauge boson upon the anomalous magnetic moment of the electron and the muon.  
The typical couplings that we will probe here will have a small effect on these quantities but not enough to either violate observational constraints nor solve the well publicised discrepancy between theory and experiment \cite{Boehm:2004uq}.

\section{Results}
\begin{figure}[!htb]
    \centering
    \includegraphics[width=.48\textwidth]{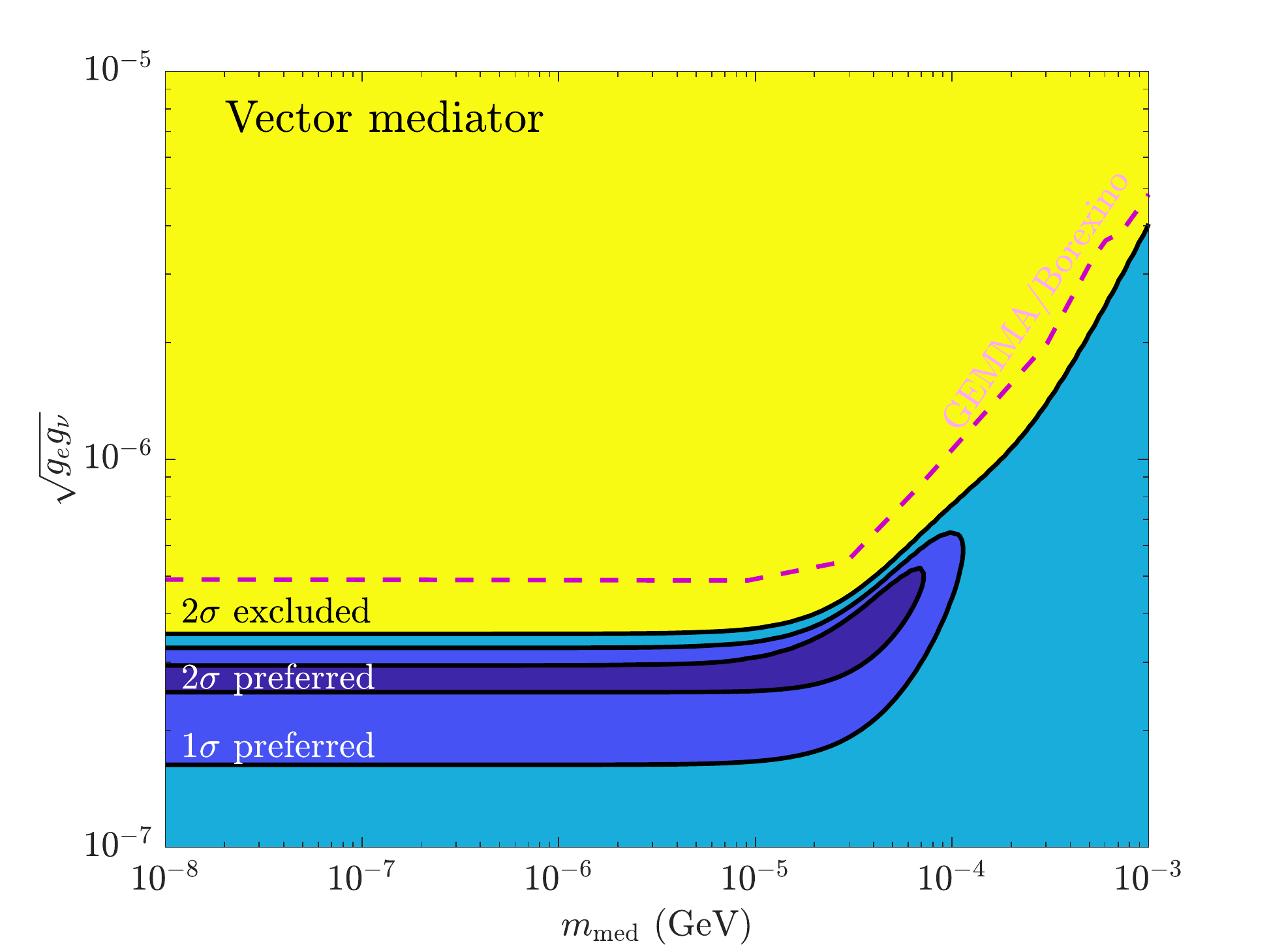}    \includegraphics[width=.48\textwidth]{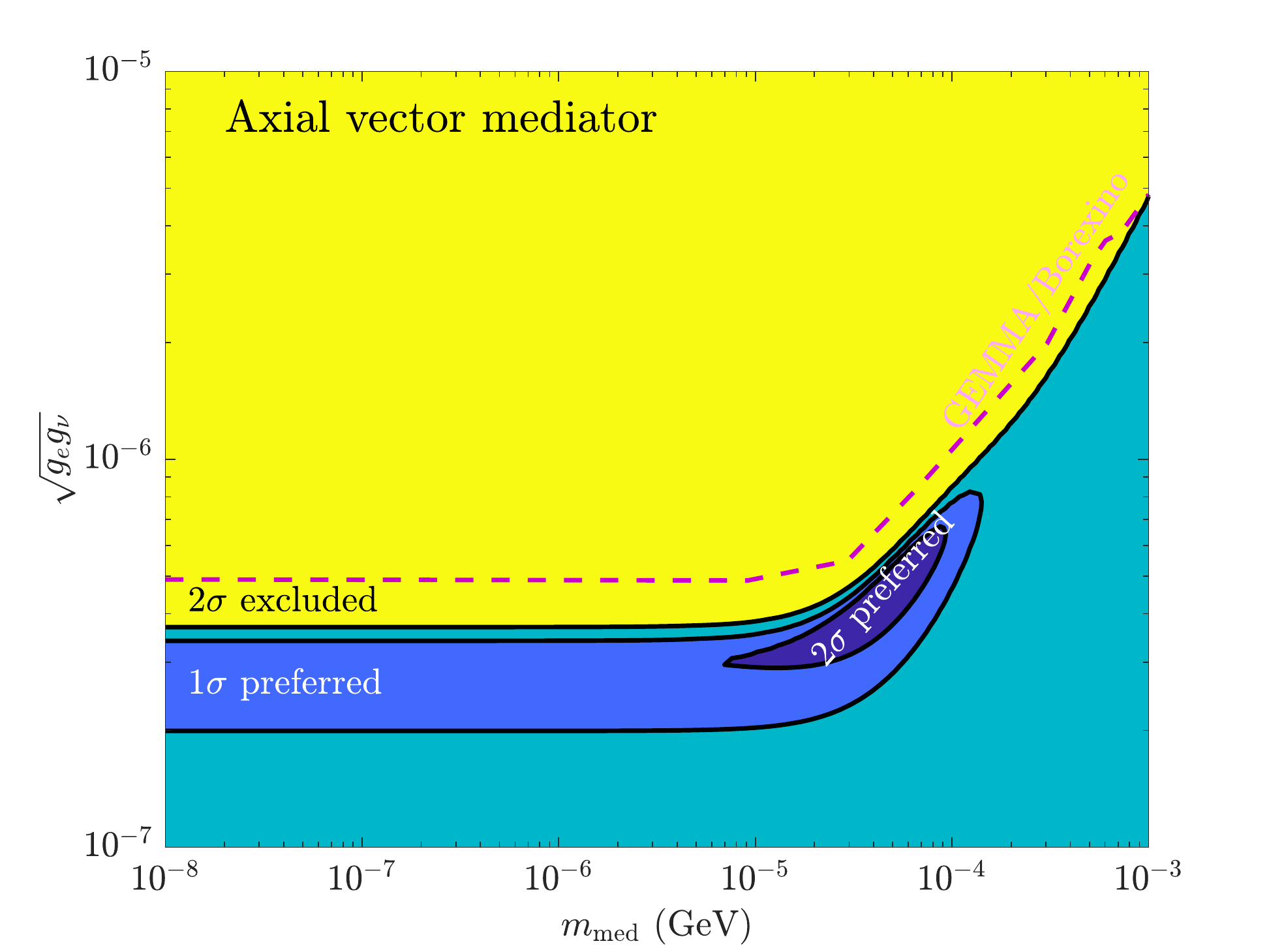}
        \includegraphics[width=.48\textwidth]{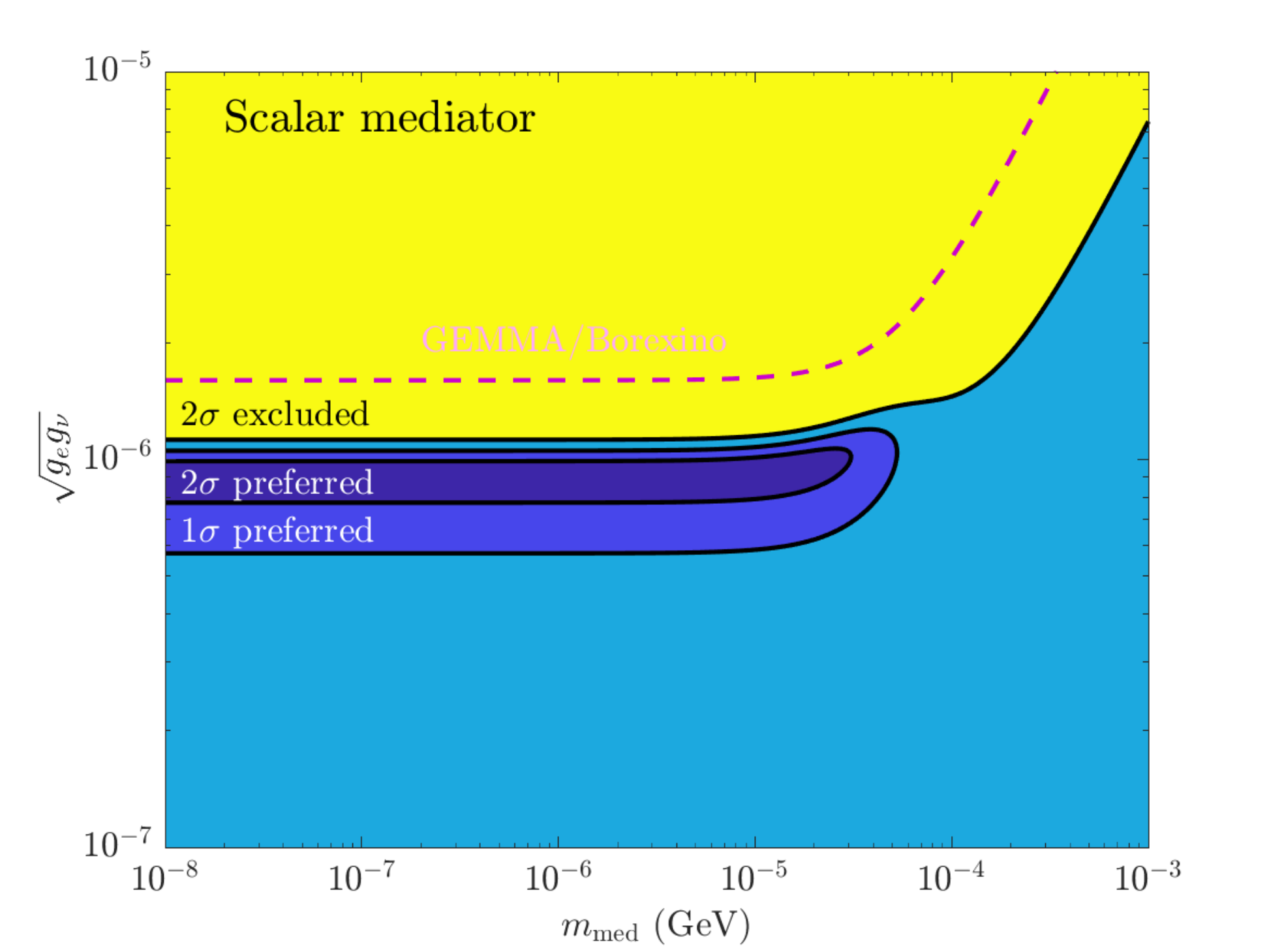}
    \caption{Constraints from the low-energy XENON1T data on the interaction of solar neutrinos via a new vector, axial vector or scalar mediator. Dark blue regions correspond to an improved fit by 1 and 2 sigma over the standard model-only hypothesis, while the yellow region is excluded by the data at 95\% CL.  In dashed pink we plot the upper limit from GEMMA \cite{Beda:2013mta}.}
    \label{fig:2dequalcouplings}
\end{figure}
In this section we show results of analysing the XENON1T electron recoil data in light of a new scalar or vector mediator. We extract the data and errors from \cite{Aprile:2020tmw}.  We allow the overall normalisation of the background to vary with a Gaussian error of 2.6\% reflecting uncertainty in the rather flat contribution of decay products from lead over the area of interest.  We allow the efficiency to vary with a Gaussian error of 3\% around the central value reported by XENON1T. The test statistic thus constructed is the binned chi-squared:
\begin{eqnarray}
\chi^2 = \sum_{i = 1}^N\frac{\left(n_i - n_i^{\rm obs}\right)^2}{\sigma_i^2} + \left(\frac{\mathcal{E}-1}{0.03}\right)^2 + \left(\frac{\mathcal{B}-1}{0.026}\right)^2,
\end{eqnarray}
where $n_i$ is the predicted number of events per bin (including efficiency and adding backgrounds), $n_i^{\rm obs}$ is the observed number, $\sigma_i$ is the error reported on each bin, and the sum runs over all $N = 29$ energy bins. $\mathcal{E}$ and $\mathcal{B}$ are the respective scalings of the efficiency and background. 

 For each interaction type considered here (scalar, vector \& axial vector), we first perform a scan over masses and couplings to extract the significance of the improvement over the background-only hypothesis. In Fig. \ref{fig:2dequalcouplings}  we show the best fit region (blue) and excluded region (yellow) as a function of the couplings $\sqrt{g_e g_\nu}$ and mediator mass. We compare each point with the maximum likelihood in the background-only hypothesis, which yields a $\chi^2 = 46.3$. Significance is estimated assuming the test statistic follows a chi-squared distribution with 2 degrees of freedom. We also overlay constraints on $\sqrt{g_e g_\nu}$ from GEMMA and Borexino (pink, dashed).
 
\begin{figure}[t]
    \centering
    \includegraphics[width=.5\textwidth]{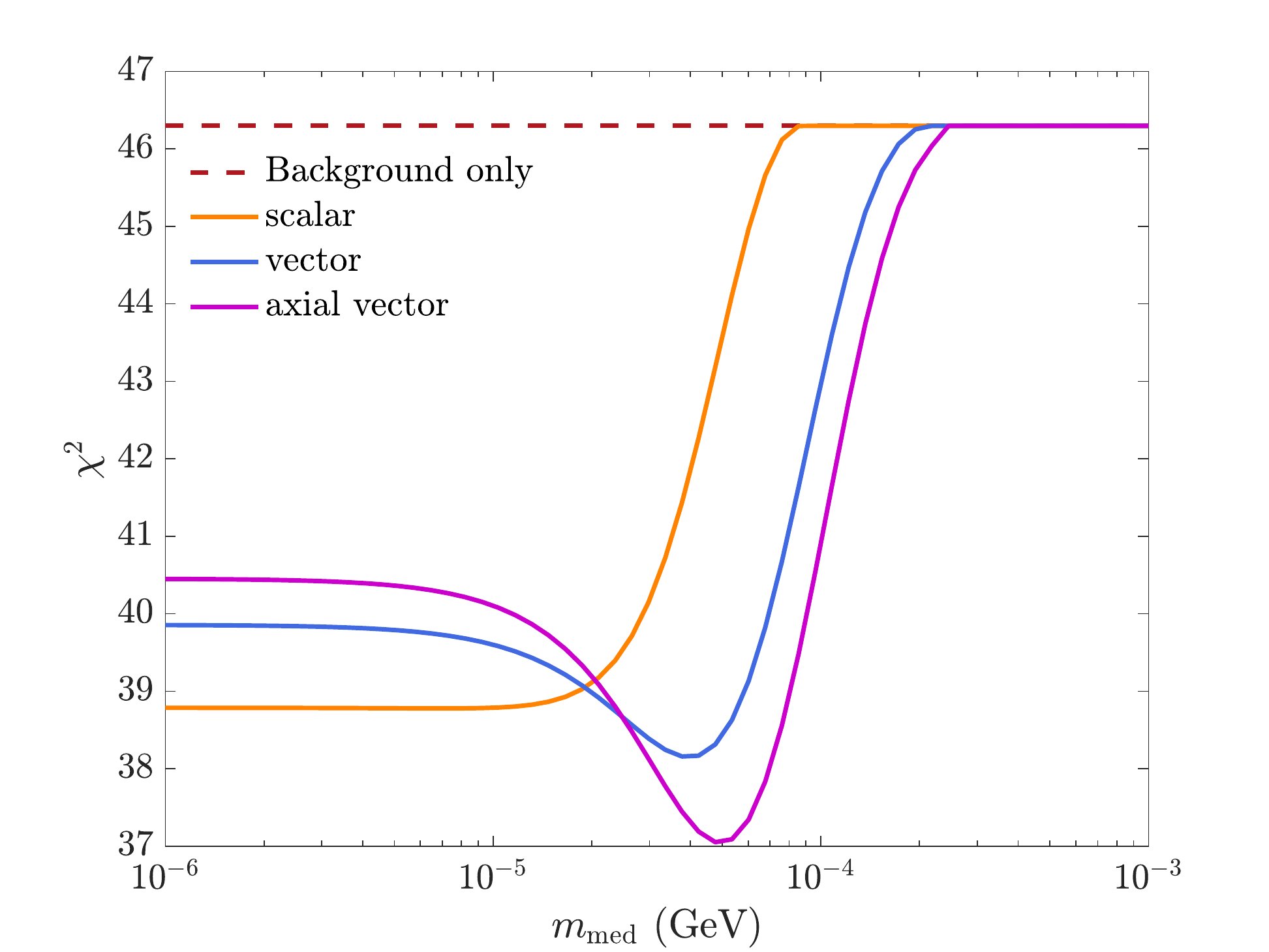}
    \caption{Minimum $\chi^2$ as a function of mediator mass for the scalar and vector mediator scenarios. The couplings of the mediator to neutrinos and electrons have been profiled over.}.
    \label{fig:chi2fig}
\end{figure}

We can see that, in both scalar and vector mediator scenarios the XENON1T excess can be explained avoiding the GEMMA/Borexino constraints.
We also note that a pure pseudoscalar yields no improvement in this parameter space.  The resulting $\chi^2$ statistics as a function of the mediator mass are shown in Fig. \ref{fig:chi2fig}, profiled over values of the coupling $\sqrt{g_e g_\nu}$. While the axial vector case gives a slightly stronger improvement over the SM, it does not quite rise to the 3$\sigma$ level for two degrees of freedom.

\begin{figure}
    \centering
    \includegraphics[width=.5\textwidth]{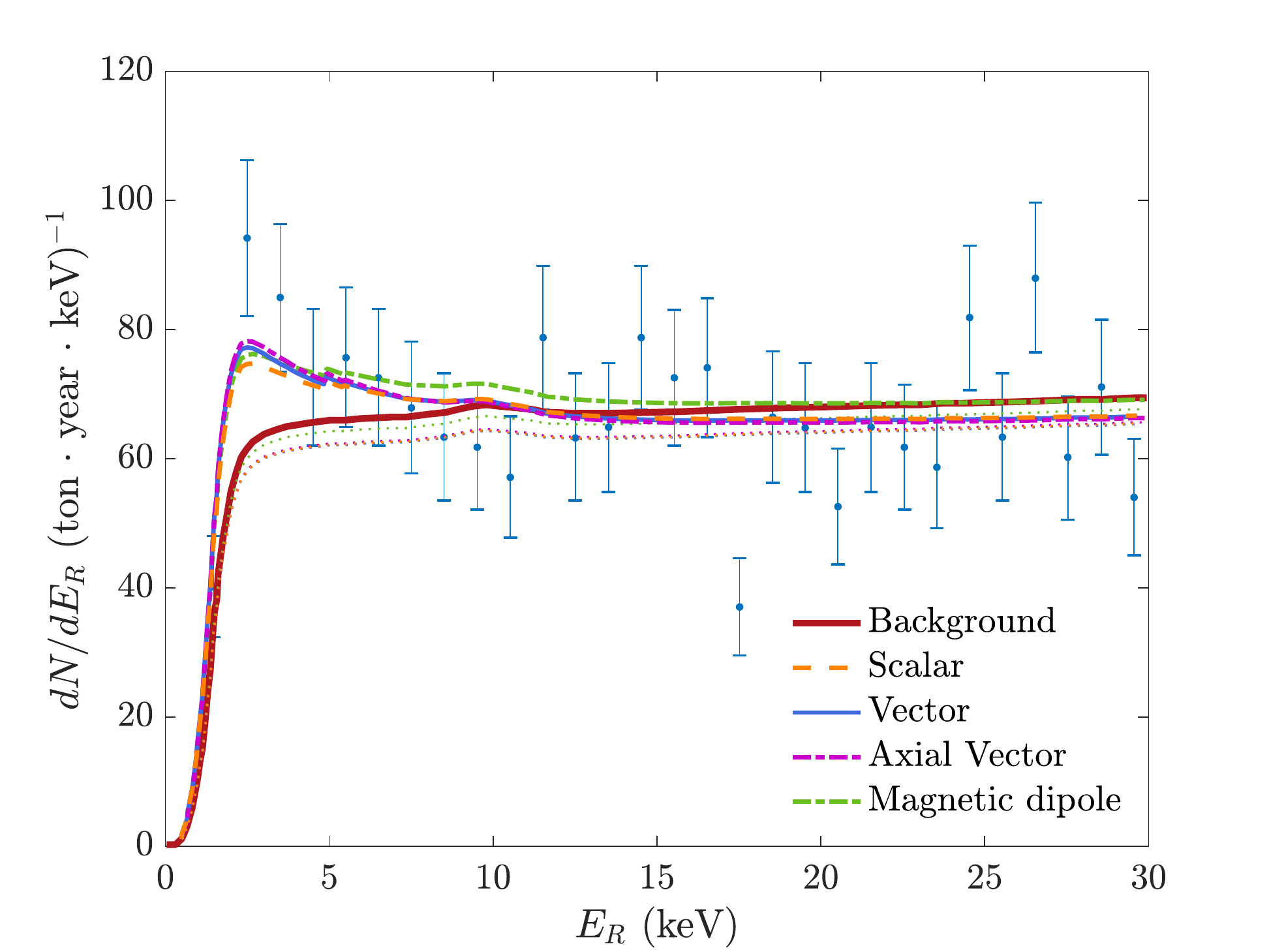}
    \caption{The low-energy electron recoil data from XENON1T, along with the best fit spectra for solar neutrinos interacting via a new  scalar (dashed orange $m_{\phi} = 9$ keV, $\sqrt{g_e g_\nu} = 9\times 10^{-7}$), or vector (blue $m_{Z'} = 41$ keV, $\sqrt{g_e g_\nu} = 3.9\times 10^{-7}$) mediator. We also show the case of a neutrino magnetic dipole moment (green, $\mu =  2.3\times 10^{-11} \mu_B$). }
    \label{fig:my_label}
\end{figure}

The electron recoil spectra expected given our best fits are plotted in Fig. \ref{fig:my_label} as solid lines. For each case, we show the best fit differential scattering rate $dN/dE_R$, adding signal and  background. 
We also show, as a dotted line, the background only fit hypothesis, taking into account systematic uncertainties and efficiencies.
Finally, we include in dashed green the best fit that we obtained using the same technique with the neutrino magnetic dipole model explored in \cite{Aprile:2020tmw}. The latter is slightly less favoured ($\Delta \chi^2 \simeq 1$) than the light mediator model, due to a higher predicted rate at energies between 5 and 15 keV leading to less of a ``peak'' at low energies.

\section{Discussion and Conclusion}
\label{sec:conclusions}
In the results section we saw that the excess observed in electron recoil events might be explained by an enhanced cross section between neutrinos and electrons which would boost the interaction of $pp$ neutrinos with the electrons in the xenon atoms making up the target mass of the XENON1T detector.  
We will discuss the vector mediator interpretation. 
Similar considerations can be made for the scalar mediator scenario.
We observed that couplings of $\sqrt{g_eg_\nu}\sim3\times 10^{-7}$ and a mediator mass $m_{\mathrm{med}}\sim 20-50$ keV provided the best fit. This increased the goodness of fit of the data to the model by $\Delta\chi^2=9.3$.  
There is a complete degeneracy between $g_e$ and $g_\nu$ in this energy range so there is no motivation for fitting the excess in allowing the two couplings to vary independently, although there might be if one is trying to evade astrophysical constraints.  

This coupling avoids the constraints from GEMMA and Borexino on enhanced interactions between neutrinos and electrons, as well as constraints from other dark matter detectors.  When we allow $g_\nu$ to vary independently of $g_e$ we find that there is effectively complete degeneracy and that the only criterion required from the couplings is that the product has the correct value.

It is only just possible to accommodate the value $\sqrt{g_eg_\nu}\sim3\times 10^{-7}$ within the scope of terrestrial experiments -- the constraint from GEMMA is around $\sqrt{g_eg_\nu}< 5\times 10^{-7}$.  If we view our fit as a maximum possible contribution of new physics, i.e. a constraint rather than an indication of beyond standard model physics, this means that the XENON1T collaboration are about to lead the field on setting constraints on new light mediator interactions between electrons and neutrinos.  This is a significant achievement and demonstrates yet another way in which large ultra low background dark matter detectors designed to search for WIMPs are able to place constraints on a myriad of new physics in different areas.

However, returning to our interpretation of the excess as being due to new physics, we do face a very significant problem in the form of constraints from astrophysics.  Couplings between neutrinos and electrons via new mediators inevitably lead to new $\nu-\nu$ and $e-e$ self-interactions which themselves create problems.  If we set the two couplings equal to each other, the best fit region is close to the boundary where energy loss in core collapse supernovae would ultimately result in fewer overall neutrinos from SN1987A \cite{Heurtier:2016otg,Farzan:2002wx}.  At the same time the $e-e$ interaction corresponding to that coupling implies enhanced energy transport in both the Sun and especially red giant stars, changing the temperature profile in the core at the onset and during helium burning, which would change the shape of both the red giant branch and the horizontal branch in the HR diagram of globular clusters \cite{Hardy:2016kme,Redondo:2013lna}.

Often such additional contributions to the energy transport inside stars provide strong constraints which stop at extremely low couplings, far below what we are considering for $g_e$ at the mass in question (20-40 keV is constrained in the Sun but is even more tightly constrained in the hotter interior of the deep cores of red giant stars) but also can {\it in principle} run out of constraining power at very high couplings. Indeed if the couplings are high enough, any new mediators which are produced on shell in one part of the star are re-absorbed soon afterwards due to a lower mean free path with respect to the temperature scale height.  In this situation, one can imagine that for very large $g_e$ couplings astrophysical constraints could therefore weaken. Finally, we note that $g_e$ couplings could be brought as high as around $3\times 10^{-5}$ before coming into tension with measurements of the electron anomalous magnetic moment $(g-2)_e$ \cite{Hanneke:2010au}.

Relatively recent in-depth studies into the precise topic of such astrophysical constraints have considered these possibilities and not found any obvious loopholes at larger couplings \cite{Hardy:2016kme}.  There are also constraints on $g_e$ from supernovae which might be even harder to avoid \cite{Chang:2018rso}.  We note that there are exotic models where the mass of mediators depends upon the local environment with varying levels of complexity (see for example \cite{Nelson:2008tn}) and that if in the future this signal became stronger and astrophysical bounds remain prohibitive at those couplings, one might have to revisit those rather complicated scenarios.

In summary, the recent result from the XENON1T analysis shows that we are entering an era where the ultra low background, ultra sensitive dark matter detectors of the current era are reaping benefits in areas of physics other than what they are designed for.  Viewed as a signal for new physics, the excess can be fit by various models including the one set out in this paper, but most of them including the one presented above
are at odds with independent astrophysical bounds.  Viewed as a constraint on new physics, it demonstrates that dark matter experiments are starting to constrain the properties of the neutrino sector.  In the next years we expect the XENONnT \cite{Aprile:2015uzo} and LZ \cite{Mount:2017qzi} experiments to start to operate with around 5 times the exposure of XENON1T which will tighten the constraints on new physics.  Future argon experiments like Darkside-20k \cite{Aalseth:2017fik} and the potentially even larger xenon experiments such as DARWIN \cite{Aalbers:2016jon,Aalbers:2020gsn} lead to exciting possibilities for probing a huge variety of physics beyond the standard model. 

\section*{acknowledgements}
We are grateful for conversations with Miguel Campos and Miguel Escudero.  MF acknowledges support from the STFC and funding from the European Research Council under the European Union's Horizon 2020 programs (ERC Grant Agreement no.648680 DARKHORIZONS). ACV is supported by the Arthur B. McDonald Canadian Astroparticle Physics Research Institute, with equipment funded by the Canada Foundation for Innovation and the Ontario Ministry of Economic Development, Job Creation and Trade (MEDJCT). Research at Perimeter Institute is supported by the Government of Canada through the Department of Innovation, Science, and Economic Development, and by the Province of Ontario through MEDJCT. DGC acknowledges financial support from the project SI2/PBG/2020-00005 and is also supported in part by the Spanish Agencia Estatal de Investigaci\'on through the grants PGC2018-095161-B-I00 and IFT Centro de Excelencia Severo Ochoa SEV-2016-0597, and the Spanish Consolider MultiDark FPA2017-90566-REDC.
This manuscript has been authored by Fermi Research Alliance, LLC under Contract No. DE-AC02-07CH11359 with the U.S. Department of Energy, Office of Science, Office of High Energy Physics.

\bibliography{biblio}

\end{document}